\documentclass[a4paper,11pt]{article}
\usepackage{pos}

\usepackage{amsmath,amsfonts,mathtools,mathrsfs,bbold,slashed}

\usepackage[normalem]{ulem}

\title{SymEFT for local tastes of staggered lattice QCD}
\author*[a,b]{Nikolai Husung}

\affiliation[a]{Instituto de Física Teórica UAM-CSIC,\\
   C/ Nicolás Cabrera 13-15, Universidad Autónoma de Madrid, Cantoblanco 28049 Madrid, Spain}
\affiliation[b]{Departamento de Física Teórica, Universidad Autónoma de Madrid,\\
   Cantoblanco 28049 Madrid, Spain}

\emailAdd{nikolai.husung@uam.es}

\abstract{The applicability of Symanzik Effective Field Theory (SymEFT) for the description of lattice artifacts assumes a local formulation of the lattice theory.
We discuss the symmetries realised by tastes local in spacetime of unrooted staggered quarks, approaching mass-degenerate 4-flavour QCD in the continuum limit.
An outlook on some implications for the asymptotic lattice-spacing dependence is given for spectral quantities as well as local composite fields.}

\FullConference{The 41st International Symposium on Lattice Field Theory (LATTICE2024)\\
 28 July - 3 August 2024\\
Liverpool, UK\\}

\def\taste{\Phi}
\def\tastebar{\bar{\Phi}}
\def\ctaste{\Psi}
\def\ctastebar{\bar{\Psi}}
\def\Nc{N}
\def\Nf{N_\mathrm{f}}
\def\Nset{N_\mathrm{sets}}
\def\nmin{n_\mathrm{min}}
\def\gbar{\bar{g}}
\def\unity{\mathbb{1}}
\def\ord{\mathop{\mathrm{O}}}
\def\L{\mathscr{L}}
\def\tr{\mathop{\mathrm{tr}}}
\def\diag{\mathop{\mathrm{diag}}}
\def\rmd{\mathrm{d}}

\def\MSbar{\overline{\mathrm{MS}}}

\newcommand{\op}{\mathcal{O}}

\newcommand{\base}{\mathcal{B}}
\newcommand{\opE}{\mathcal{E}}

\newcommand{\opFive}{\op^{(1)}}
\newcommand{\opSix}{\op^{(2)}}

\newcommand{\opFiveE}[1][]{\opFive_{\opE #1}}

\newcommand{\cev}[1]{\overset{\leftarrow}{#1}}

\def\Dslash{\slashed{D}}

\begin{document}
\maketitle

\section{Introduction}
One of the central ingredients to extract continuum physics from lattice QCD simulations is the continuum extrapolation of renormalised quantities in the lattice spacing $a\searrow0$.
Reaching the continuum limit with sufficient control over the associated systematic errors requires knowledge about the asymptotic lattice spacing dependence of the quantity of interest.

Due to working in an asymptotically free theory, we do know that the asymptotically leading lattice spacing dependence is of the form $a^{\nmin}[2b_0\gbar^2(1/a)]^{\hat{\Gamma}_i}$, where $48\pi^2b_0=11\Nc-2\Nf$ with $\Nc$ colours as well as $\Nf$ flavours, and $\gbar(1/a)$ is the running coupling at renormalisation scale $\mu=1/a$ corresponding to the relevant scale of lattice artifacts.
A priori there are no bounds on the values allowed for $\hat{\Gamma}_i$ and they may become severely negative thus worsening the approach to the continuum limit.
That this is not purely an academical concern, can be seen in the seminal work by Balog, Niedermayer and Weisz~\cite{Balog:2009yj,Balog:2009np} in the O(3) non-linear sigma model, where they found $\min_i\hat{\Gamma}_i=-3$.

To check for any such issues in lattice QCD one needs to derive these powers for the lattice discretisation of choice in terms of a Symanzik Effective Field Theory~(SymEFT) \cite{Symanzik:1979ph,Symanzik:1981hc,Symanzik:1983dc,Symanzik:1983gh} analysis.
Here the lattice artifacts of the renormalised theory are treated as a perturbation around the continuum limit, i.e., we can formally write the SymEFT in terms of the effective Lagrangian
\begin{equation}
\L_\mathrm{eff}=\L_\mathrm{QCD}+a\L^{(1)}+a^2\L^{(2)}+\ord(a^3)\,,
\end{equation}
where $\L^{(d)}$ is a linear combination of operators of mass-dimension $(4+d)$ that comply with the symmetry constraints of the lattice action.
Eventually one finds as many $\hat{\Gamma}_i$ at a given power in the lattice spacing as there are linearly-independent operators in the minimal basis.
The Lagrangian of the continuum theory, here Quantum Chromodynamics~(QCD), reads
\begin{equation}
\L_\mathrm{QCD}=-\frac{1}{2g_0^2}\tr(F_{\mu\nu}F_{\mu\nu})+\bar{\Psi}\left\{\gamma_\mu D_\mu+M\right\}\Psi\,,
\end{equation}
where $F_{\mu\nu}=[D_\mu,D_\nu]$ is the field-strength tensor and $D_\mu=\partial_\mu+A_\mu$ is the covariant derivative with algebra-valued gauge field $A_\mu\in\mathrm{su}(\Nc)$.
The \emph{continuum} quark-flavours $\Psi$ are here assumed to be 4-fold mass-degenerate, i.e., $M=\diag(m_1,\ldots,m_{\Nset})\otimes\unity_{4\times 4}$ with $\Nset < 11\Nc/8$.
This particular choice is due to discussing lattice QCD with \emph{unrooted} Staggered quarks as initially proposed by Kogut and Susskind~\cite{Kogut:1974ag}, which by construction yield (multiples of) four mass-degenerate flavours in the continuum limit, i.e., $\Nf=4\Nset$.

The Staggered lattice fermion action reads in the one-component representation~(1CR)
\begin{equation}
S_\mathrm{1CR}=a^4\sum_x\bar{\chi}(x)\frac{\eta_\mu(x)}{2}\left\{\nabla_\mu+\nabla_\mu^*\right\}\chi(x),\quad \eta_\mu(x)=(-1)^{\sum_{\nu<\mu}x_\nu/a},
\end{equation}
where $\chi$ are the quark fields in the 1CR and we introduced the covariant forward and backward lattice derivatives
\begin{equation}
a\nabla_\mu \chi(x)=U(x,\mu)\chi(x+a\hat{\mu})-\chi(x),\quad a\nabla_\mu^* \chi(x)=\chi(x)-U^\dagger(x-a\hat{\mu},\mu)\chi(x-a\hat{\mu}),
\end{equation}
with gauge links $U(x,\mu)\in\text{SU}(\Nc)$.

To work out the connection to the continuum notion of quark flavours, we need to identify the four \emph{tastes} contained within the 1CR, which will become the flavours of the continuum theory.
Identifying those tastes is typically referred to as a taste representation~(TR).
The most commonly used TR are momentum-space tastes, which use the 16 corners of the Fourier-space hypercube to construct tastes.
Unfortunately, the resulting TR suffers from non-local interactions~\cite{Kluberg-Stern:1983lmr,Daniel:1986zm,Jolicoeur:1986ek}, which violates one of the central assumptions made when constructing a \emph{local} effective Lagrangian as we want to do in section~\ref{sec:SymEFT}.
Instead we choose a TR that is strictly local by joining the 16 corners of a spacetime hypercube, see e.g.~\cite{Kluberg-Stern:1983lmr,Verstegen:1985kt},
\begin{equation}
\tastebar(y)=\frac{1}{8}\,\,\sum_{\mathclap{\xi\in\{0,1\}^4}}\bar\chi(y+a\xi)T[U](y,\xi)^\dagger,\quad\taste(y)=\frac{1}{8}\,\,\sum_{\mathclap{\xi\in\{0,1\}^4}}T[U](y,\xi)\chi(y+a\xi),\quad y\in 2a\mathbb{Z}^4,\label{eq:localTastes}
\end{equation}
where we choose ($\mu$ increasing towards the right-hand side)
\begin{equation}
T[U](y,\xi)=\prod_{\mu=0}^3U^{\xi_\mu}(y+a{\textstyle\sum_{\nu<\mu}\xi_\nu\hat{\nu}},\mu)\gamma_\mu^{\xi_\mu}.
\end{equation}
The resulting matrix $\Phi(y)$ has one colour index and two Dirac indices $(i,f)$, where the latter are identified as spin $i$ and taste $f$.
There is quite some freedom in choosing a gauge-covariant $T[U](y,\xi)$ beyond the requirement $TT^\dagger\propto \mathbb{1}$ in order to keep the measure unchanged.
Different choices must yield equivalent TRs as they are related by a substitution in the path-integral if they keep the measure invariant.

\section{Symmetries of local tastes}
Expressing the lattice action in terms of the strictly local tastes becomes a highly non-trivial expression in the interacting theory and we omit it entirely.
The symmetries in this TR can nonetheless be inferred from their 1CR counterparts by use of eq.~\eqref{eq:localTastes}.
This is needed in order to understand what remains of the continuum symmetries and make the connection to the continuum flavours present in the SymEFT.
The symmetries found for local tastes are
\begin{subequations}\label{eq:SymmetryTransfs}
\begin{itemize}
\item SU($\Nc$) gauge symmetry,
\item U(1)${}_\mathrm{B}$ flavour symmetry,
\item Remnant of chiral symmetry.
Analogously to conventional chiral symmetry we may introduce the shorthands
\begin{align}
\tastebar_\mathrm{R}&=\tastebar\frac{1-\gamma_5\otimes\tau_5}{2}\,,&\taste_\mathrm{R}&=\frac{1+\gamma_5\otimes\tau_5}{2}\taste\,,\nonumber\\\tastebar_\mathrm{L}&=\tastebar\frac{1+\gamma_5\otimes\tau_5}{2}\,,&
\taste_\mathrm{L}&=\frac{1-\gamma_5\otimes\tau_5}{2}\taste\,,
\end{align}
where $\tau_\mu=\gamma_\mu^T$ acts in taste-space.
The massless lattice action written in this form is then invariant under
\begin{align}
\tastebar_\mathrm{R}&\rightarrow\tastebar_\mathrm{R}e^{i\vartheta_\mathrm{R}\gamma_5\otimes \tau_5-i\varphi_\mathrm{R}},& \taste_\mathrm{R}&\rightarrow e^{i\vartheta_\mathrm{R}\gamma_5\otimes \tau_5+i\varphi_\mathrm{R}}\taste_\mathrm{R},\nonumber\\\tastebar_\mathrm{L}&\rightarrow\tastebar_\mathrm{L}e^{i\vartheta_\mathrm{L}\gamma_5\otimes \tau_5-i\varphi_\mathrm{L}},&\taste_\mathrm{L}&\rightarrow e^{i\vartheta_\mathrm{L}\gamma_5\otimes \tau_5+i\varphi_\mathrm{L}}\taste_\mathrm{L},
&\vartheta_\mathrm{L,R},\varphi_\mathrm{L,R}&\in\mathbb{R}.\label{eq:LRsymm}
\end{align}
\item \emph{Modified} charge conjugation
\begin{align}
\tastebar(y)&\rightarrow-\taste^T(y)C\otimes (C^{-1})^T,\quad \taste(y)\rightarrow C^{-1}\otimes C^T\tastebar^T(y),\nonumber\\
U_\mu(x)&\rightarrow U_\mu^*(x),\quad C\gamma_\mu C^{-1}=-\gamma_\mu^T.
\end{align}
\item[$\circ$] \emph{Modified} Euclidean reflections~\cite{Verstegen:1985kt} in $\hat{\mu}$ direction
\begin{align}
\tastebar(y)&\rightarrow \tastebar(y-2y_\mu\hat{\mu})\gamma_5\gamma_\mu\otimes \tau_5\left\{1+a^2\ldots\right\},\nonumber\\
\taste(y)&\rightarrow\gamma_\mu\gamma_5\otimes\tau_5\left\{1+a^2\ldots\right\}\taste(y-2y_\mu\hat{\mu})\,,\nonumber\\
U_\nu(x)&\rightarrow\begin{cases}
U_\mu^\dagger(x-2x_\mu\hat{\mu}) & \text{if $\mu=\nu$} \\
U_\nu(x-2x_\mu\hat{\mu}) & \text{else}
\end{cases}.\label{eq:modReflections}
\end{align}
\item[$\circ$] \emph{Modified} discrete rotations~\cite{Mitra:1983bi,Verstegen:1985kt} of $90^\circ$ in any $\rho$-$\sigma$-plane, i.e., $\rho<\sigma$,
\begin{align}
\tastebar(y)&\rightarrow\frac{1}{2}\tastebar(R^{-1}y)(\unity +\gamma_\rho\gamma_\sigma)\otimes (\tau_\rho-\tau_\sigma),\nonumber\\
\taste(y)&\rightarrow\frac{1}{2}(\unity -\gamma_\rho\gamma_\sigma)\otimes (\tau_\rho-\tau_\sigma)\taste(R^{-1}y),\nonumber\\
U_\nu(x)&\rightarrow\begin{cases}
U_\rho^\dagger(R^{-1}x-a\hat{\rho}) & \nu=\sigma\\
U_\sigma(R^{-1}x) & \nu=\rho\\
U_\nu(R^{-1}x) & \text{else}
\end{cases}\nonumber\\
(R^{-1}x)_\rho&=x_\sigma,\quad (R^{-1}x)_\sigma=-x_\rho,\quad (R^{-1}x)_{\mu\neq \rho,\sigma}=x_\mu,
\end{align}
with rotation matrix $R$ acting on vectors in Euclidean spacetime.
\item[$\circ$] Shift-symmetry by a single lattice spacing in direction $\hat{\mu}$ combined with a discrete flavour rotation and field redefinition
\begin{align}
\bar\taste(y)&\rightarrow \bar\taste(y)\unity\otimes\tau_\mu \left\{1+2a\hat{P}_-^{(\mu)}\hat{\overline{\nabla}}{}_\mu^{\dagger}+a^2\ldots\right\},\nonumber\\
\taste(y)&\rightarrow \unity\otimes\tau_\mu \left\{1+2a\hat{P}_+^{(\mu)}\hat{\overline{\nabla}}_\mu+a^2\ldots\right\}\taste(y),
\end{align}
where $\hat{\overline{\nabla}}_\mu$ involves a $\hat\mu$-dependent fat link spanning a distance of $2a$ and we introduced the projectors
\begin{equation}
\hat{P}_\pm^{(\mu)}=\frac{1\pm\gamma_\mu\gamma_5\otimes\tau_\mu\tau_5}{2}.
\end{equation}
\end{itemize}
\end{subequations}
Filled ($\bullet$) or open ($\circ$) symbols highlight symmetries that are in their canonical form as one would expect in a continuum theory and those that require field-redefinitions respectively.
The latter symmetries are only written out explicitly up to $\ord(a^2)$ terms.
In the free theory, only Shift-symmetry requires a field-redefinition which further simplifies to being just a forward lattice derivative.

\section{Symanzik Effective Field Theory}\label{sec:SymEFT}
Before we can impose these symmetries we need to work out how to deal with the field-redefinitions present in some of the symmetry transformations on the lattice.
As it turns out, such a field-redefinition gives rise to operators vanishing by the fermion EOMs in the SymEFT that are violating the symmetry in its canonical form without the field-redefinition.
These particular operators start to appear at the order in the lattice spacing at which the field-redefinition arise, e.g., for Shift-symmetry we find at $\ord(a)$
\begin{align}
\opFiveE[;1]&=\frac{1}{2}\ctastebar\left\{\cev{\Dslash}\cev{D}_\nu\gamma_\nu\gamma_5\otimes\tau_\nu+\gamma_\nu\gamma_5\otimes \tau_\nu D_\nu\Dslash\right\}\ctaste,\nonumber\\
\opFiveE[;2]&=\frac{1}{2}\ctastebar\left\{\cev{\Dslash}{}^2\gamma_5\otimes\tau_\nu-\gamma_5\otimes\tau_\nu\Dslash^2\right\}\ctaste,\label{eq:opFive}
\end{align}
where we introduce the sloppy shorthands for the fermion EOMs
\begin{equation}
\ctastebar\cev{\Dslash}=\ctastebar (\gamma_\mu \cev{D}_\mu-M)\stackrel{\text{EOM}}{=}0,\quad \Dslash\ctaste=(\gamma_\mu D_\mu+M)\ctaste\stackrel{\text{EOM}}{=}0.
\end{equation}
By induction one can work out that those field-redefinitions should be curable (perturbatively) in the SymEFT, thus restoring the canonical form of the symmetry transformations to all orders in the lattice spacing.

Common lore allows us to make use of the continuum EOMs\footnote{Of course this also includes the gluon EOM
\begin{equation*}
[D_\nu,F_{\nu\mu}]=T^ag_0^2\ctastebar\gamma_\mu T^a\ctaste.
\end{equation*}} to work out the minimal \emph{on-shell} operator basis~\cite{Luscher:1996sc}.
However, ``making use of the EOMs'' eventually amounts to an overall field-redefinition (or more naturally a change of matching condition) that can and will have an impact on the matching coefficients of the SymEFT action at subleading order in the lattice spacing, here $\ord(a^2)$, as well as the matching coefficients of local composite fields at the current order, here already $\ord(a)$.
For more details on this, see~\cite{Capitani:1999ay,Capitani:2000xi,Husung:localFields}.

Since the symmetries \emph{only} permit EOM-vanishing operators in the SymEFT at $\ord(a)$, the leading-order minimal on-shell basis consists of mass-dimension~6 operators that contribute at $\ord(a^2)$.
Taking all the symmetries into account leaves us with the minimal on-shell basis
\begingroup\allowdisplaybreaks
\begin{align}
\opSix_{1}&=\frac{1}{g_0^2}\tr([D_\mu, F_{\nu\rho}]\,[D_\mu, F_{\nu\rho}])
\,,&
\opSix_{2}&=\frac{1}{g_0^2}\sum\limits_{\mu}\tr([D_\mu, F_{\mu\nu}]\,[D_\mu, F_{\mu\nu}])\,,&\nonumber\\
\opSix_{3}&=\sum_\mu\bar\ctaste\gamma_\mu\otimes\unity  D_\mu^3\ctaste,&
\opSix_{4}&=g_0^2(\bar\ctaste\gamma_\mu\otimes\unity \ctaste)^2,\nonumber\\
\opSix_{5}&=g_0^2(\bar\ctaste\gamma_\mu\gamma_5\otimes\unity \ctaste)^2,&
\opSix_{6}&=g_0^2(\bar\ctaste\gamma_\mu\otimes\unity  T^a\ctaste)^2,\nonumber\\
\opSix_{7}&=g_0^2(\bar\ctaste\gamma_\mu\gamma_5\otimes\unity T^a\ctaste)^2,&
\opSix_{8}&=\frac{i}{4}\ctastebar M\gamma_{\mu\nu}\otimes\unity F_{\mu\nu}\ctaste,\nonumber\\
\opSix_{9}&=\ctastebar M^3\ctaste,&
\opSix_{10}&=\tr(M^2)\ctastebar M\ctaste,\nonumber\\
\opSix_{11}&=g_0^2(\ctastebar\unity \otimes \tau_\mu\ctaste)^2,&
\opSix_{12}&=g_0^2(\ctastebar\unity \otimes \tau_\mu\tau_5\ctaste)^2,\nonumber\\
\opSix_{13}&=g_0^2(\ctastebar\unity \otimes \tau_\mu T^a\ctaste)^2,&
\opSix_{14}&=g_0^2(\ctastebar\unity \otimes \tau_\mu\tau_5 T^a\ctaste)^2,\nonumber\\
\opSix_{15}&=g_0^2(\ctastebar\gamma_5\otimes \tau_\mu \ctaste)^2,&
\opSix_{16}&=g_0^2(\ctastebar\gamma_5\otimes \tau_\mu\tau_5 \ctaste)^2,\nonumber\\
\opSix_{17}&=g_0^2(\ctastebar\gamma_5\otimes \tau_\mu T^a\ctaste)^2,&
\opSix_{18}&=g_0^2(\ctastebar\gamma_5\otimes \tau_\mu\tau_5 T^a\ctaste)^2,\nonumber\\
\opSix_{19}&=g_0^2(\ctastebar\gamma_\mu\otimes \tau_5 \ctaste)^2,&
\opSix_{20}&=g_0^2(\ctastebar\gamma_\mu\gamma_5\otimes \tau_5 \ctaste)^2,\nonumber\\
\opSix_{21}&=g_0^2(\ctastebar\gamma_\mu\otimes \tau_5 T^a\ctaste)^2,&
\opSix_{22}&=g_0^2(\ctastebar\gamma_\mu\gamma_5\otimes \tau_5 T^a\ctaste)^2,\nonumber\\
\opSix_{23}&=g_0^2(\ctastebar\gamma_{\mu\nu}\otimes \tau_\rho \ctaste)^2,&
\opSix_{24}&=g_0^2(\ctastebar\gamma_{\mu\nu}\otimes \tau_\rho\tau_5 \ctaste)^2,\nonumber\\
\opSix_{25}&=g_0^2(\ctastebar\gamma_{\mu\nu}\otimes \tau_\rho T^a\ctaste)^2,&
\opSix_{26}&=g_0^2(\ctastebar\gamma_{\mu\nu}\otimes \tau_\rho\tau_5 T^a\ctaste)^2,\nonumber\\
\opSix_{27}&=g_0^2(\ctastebar\gamma_\mu\otimes \tau_{\nu\rho} \ctaste)^2,&
\opSix_{28}&=g_0^2(\ctastebar\gamma_\mu\gamma_5\otimes \tau_{\nu\rho} \ctaste)^2,\nonumber\\
\opSix_{29}&=g_0^2(\ctastebar\gamma_\mu\otimes \tau_{\nu\rho} T^a\ctaste)^2,&
\opSix_{30}&=g_0^2(\ctastebar\gamma_\mu\gamma_5\otimes \tau_{\nu\rho} T^a\ctaste)^2,\nonumber\\
\opSix_{31}&=g_0^2\sum_\mu(\ctastebar\gamma_{\mu\nu}\otimes \tau_\mu \ctaste)^2,&
\opSix_{32}&=g_0^2\sum_\mu(\ctastebar\gamma_{\mu\nu}\otimes \tau_\mu\tau_5 \ctaste)^2,\nonumber\\
\opSix_{33}&=g_0^2\sum_\mu(\ctastebar\gamma_{\mu\nu}\otimes \tau_\mu T^a\ctaste)^2,&
\opSix_{34}&=g_0^2\sum_\mu(\ctastebar\gamma_{\mu\nu}\otimes \tau_\mu\tau_5 T^a\ctaste)^2,\nonumber\\
\opSix_{35}&=g_0^2\sum_\mu(\ctastebar\gamma_\mu\otimes \tau_{\mu\nu} \ctaste)^2,&
\opSix_{36}&=g_0^2\sum_\mu(\ctastebar\gamma_\mu\gamma_5\otimes \tau_{\mu\nu} \ctaste)^2,\nonumber\\
\opSix_{37}&=g_0^2\sum_\mu(\ctastebar\gamma_\mu\otimes \tau_{\mu\nu} T^a\ctaste)^2,&
\opSix_{38}&=g_0^2\sum_\mu(\ctastebar\gamma_\mu\gamma_5\otimes \tau_{\mu\nu} T^a\ctaste)^2.
\end{align}
\endgroup
The operators $\op_{i\leq10}$ are compatible with chiral symmetry or its spurion counterpart in the massive theory and thus sufficient for a chirally-symmetric fermion action like Ginsparg-Wilson quarks~\cite{Ginsparg:1981bj}, while the other operators in the basis account for the flavour-changing interactions allowed for Staggered quarks at $\ord(a^2)$.

Allowing for multiple sets of tastes ($\Nset>1$) requires the replacement $\tau\rightarrow \diag(\tau,\ldots,\tau)$ since any shift on the lattice has to be done for all tastes simultaneously while the four tastes in each set are connected via Shift-symmetry, remnant chiral, \emph{modified} Euclidean reflections, and \emph{modified} rotations as listed in equations~\eqref{eq:SymmetryTransfs}.
In the case of full mass-degeneracy (in the continuum limit) we may still rotate among the $n$th flavour of each set, i.e., there is an $\text{SU}(\Nset)^4$ flavour symmetry remaining from the full $\text{SU}(\Nf)$ flavour symmetry of continuum QCD.

\section{1-loop anomalous dimensions}
Eventually we are looking for the powers $[2b_0\gbar^2(1/a)]^{\hat{\Gamma}_i}$ multiplying the pure $a^2$ corrections.
Those powers can be inferred from the 1-loop anomalous dimension matrix of the full set of operators
\begin{equation}
\mu\frac{\rmd \op_{j;\MSbar}}{\rmd\mu}=-\gbar^2(\mu)\big[\gamma_0^\op+\ord(\gbar^2)\big]_{jk}\op_{k;\MSbar}\,.
\end{equation}
Through an appropriate change of operator basis $\op\rightarrow\base$ we can bring the 1-loop anomalous dimension matrix into Jordan normal form.
The on-diagonal entries
\begin{equation}
\hat{\gamma}_i=\frac{(\gamma_0^\base)_{ii}}{2b_0}
\end{equation}
then yield the desired powers in the running coupling
\begin{equation}
\hat{\Gamma}_i=\hat{\gamma}_i+n_i\label{eq:GammahatDef}
\end{equation}
up to potential suppression from vanishing matching coefficients $n_i\in\mathbb{N}\cup\{0\}$.
If the 1-loop anomalous dimension matrix is non-diagonalisable any off-diagonal entries remaining will give rise to additional $\log(\gbar(\mu))$ factors.
Here this is only relevant for $\Nf=0$.
For more details on these logarithms have a look at the discussion in~\cite{Husung:2022kvi}.

The 1-loop anomalous dimension matrix can be obtained from the 1-loop renormalisation of the operator basis, where we work in background field gauge~\cite{DeWitt:1967ub,tHooft:1975uxh,Abbott:1980hw,Abbott:1981ke} and within the $\MSbar$ renormalisation scheme~\cite{tHooft:1972tcz,tHooft:1973mfk,Bardeen:1978yd}.
The full setup is identical to our earlier work on Wilson and GW quarks~\cite{Husung:2022kvi}.
Here we only give the powers $\hat{\gamma}_i$, see table~\ref{tab:gammahat}, without taking matching into account as this will depend on the particular choice of action.
Notice that 4-quark operators violating chiral symmetry are expected to be absent at tree-level as they cannot be generated from mixing with any of the chirally-symmetric operators.
For the full mixing matrix for our initial basis see the actual paper explaining everything in more detail~\cite{Husung:staggered}.

\begin{table}\centering
\caption{List of \emph{unique} asymptotically leading powers $\hat{\gamma}_i$ found at $\ord(a^2)$ rounded to the third decimal \emph{without} taking suppression due to vanishing matching coefficients or any subleading corrections into account.
The latter will obviously start at $\hat{\gamma}_i+1$.
\dotuline{Underdotted} numbers belong to explicitly mass-dependent contributions, \underline{underlined} numbers originate from operators compatible with chiral symmetry, \dashuline{underdashed} numbers originate from operators only compatible with the reduced Staggered symmetries as detailed in equations~\eqref{eq:SymmetryTransfs}, and the \textbf{bold} numbers highlight an explicit contribution of the form $a^2[2b_0\gbar^2(1/a)]^{\hat{\gamma}_i}\log(2b_0\gbar^2(1/a))$ originating from the extended operator basis of Staggered quarks.
Powers with various markings originate from all the corresponding sets of operators.}\label{tab:gammahat}
\begin{tabular}{c|l}
$\Nf$ & $\hat{\gamma}_i$ \\\hline\hline
0  & \dotuline{$-0.273$}, \dashuline{$0.014$}, \underline{\dashuline{\boldmath$0.273$}}, \dotuline{$0.424$}, \dashuline{$0.560$}, \underline{$0.597$}, \underline{$0.634$}, \underline{\dashuline{$0.636$}}, \dotuline{$0.727$}, \dashuline{$0.939$}, \dashuline{$0.955$}, \dashuline{$0.970$}, \\
   & \underline{\dashuline{\boldmath$1.091$}}, \dashuline{$1.121$}, \underline{$1.145$}, \dashuline{$1.182$}, \underline{$1.201$}, \dashuline{$1.242$}, \dashuline{$1.501$} \\[3pt]\hline
4  & \dashuline{$-0.301$}, \dotuline{$-0.040$}, \dashuline{$0.040$}, \underline{$0.209$}, \dashuline{$0.419$}, \dashuline{$0.520$}, \dotuline{$0.560$}, \underline{$0.698$}, \underline{$0.817$}, \dashuline{$0.920$}, \dashuline{$0.941$}, \dotuline{\dashuline{$0.960$}},  \\
   & \dashuline{$1.120$}, \underline{$1.140$}, \underline{$1.160$}, \dashuline{$1.240$}, \dashuline{$1.320$}, \underline{$1.487$}, \dashuline{$1.661$}, \underline{$1.852$} \\[3pt]\hline
8  & \dashuline{$-0.913$}, \dashuline{$-0.412$}, \underline{$-0.103$}, \dashuline{$0.146$}, \dashuline{$0.294$}, \dotuline{$0.412$}, \underline{$0.770$}, \dotuline{$0.824$}, \dashuline{$0.882$}, \dashuline{$0.913$}, \dashuline{$0.941$}, \dashuline{$1.176$}, \\
   & \underline{$1.186$}, \underline{$1.218$}, \dashuline{$1.235$}, \dashuline{$1.353$}, \dotuline{$1.412$}, \dashuline{$1.471$}, \dashuline{$1.972$}, \underline{$2.176$}, \underline{$2.370$}, \underline{$3.210$} \\[3pt]\hline
12 & 
\dashuline{$-2.614$}, \dashuline{$-1.667$}, \underline{$-1.040$}, \dashuline{$-0.614$}, \dashuline{$-0.333$}, \underline{$0.731$}, \dashuline{$0.778$}, \dashuline{$0.836$}, \dashuline{$0.889$}, \dashuline{$1.333$}, \underline{$1.424$}, \dashuline{$1.444$}, \\
   & \dotuline{$1.556$}, \dotuline{\dashuline{$1.667$}}, \dashuline{$1.889$}, \underline{$2.223$}, \dotuline{$2.667$}, \dashuline{$2.836$}, \underline{$5.000$}, \underline{$5.145$}, \underline{$6.970$}
\end{tabular}
\end{table}

\section{Conclusion}
Contrary to other works on Staggered quarks we have chosen a strictly-local TR to establish the connection to SymEFT.
While this complicates the discussion of the symmetries significantly it puts the SymEFT treatment on more solid grounds as one requires a local effective Lagrangian.
Also in this setup one is able to show absence of $\ord(a)$ terms in the \emph{on-shell} minimal basis, settling the question repeatedly asked in the past whether there are such terms, see e.g.~\cite{Kluberg-Stern:1983lmr,Daniel:1986zm,Luo:1996vt,PerezRubioThesis,Lepage:2011vr}.

The operator basis found here is larger than the (different) ones found in either~\cite{Luo:1996vt,Lee:1999zxa} but agrees with the one indicated in~\cite{Follana:2006rc}.
The derivation of the basis has been done in two independent manners, where the second time an automation was devised, without any change for the minimal basis.
The renormalisation of the minimal basis further strengthens the expectation that this is indeed the minimal basis because no redundancies in the available counterterms were observed.
Of course, this only guarantees absence of overcompleteness rather than completeness of the basis.

For this basis we have calculated the powers $\hat{\gamma}_i$ relevant for the asymptotically leading corrections to classical $a^2$-scaling for spectral quantities of unrooted Staggered quarks.
Those corrections then lead to the asymptotic form $a^2[2b_0\gbar^2(1/a)]^{\hat{\Gamma}_i}$, where $\hat{\Gamma}_i\geq \hat{\gamma}_i$ accounts for potential suppression at the level of the matching coefficients, see eq.~\eqref{eq:GammahatDef}.
Lattice artifacts involving local fields such as, e.g., matrix elements and correlators get additional contributions due to the discretisation of the local fields but the corrections computed here will still contribute.
In general the severely enlarged operator basis compared to Wilson and GW quarks makes the situation even more complicated due to the various terms contributing.

Contrary to the O(3) non-linear sigma model we find $\min_i\hat{\Gamma}_i\gtrsim -0.273, -0.301$ for $\Nf=0,4$ respectively which is good news.
Meanwhile the powers for $\Nf=8,12$ start to become troublingly negative with $\min_i\hat{\Gamma}_i\gtrsim-0.913, -2.614$ respectively.
Here one has to remember that the chiral-symmetry breaking 4-quark operators, which are responsible for those powers, are expected to be suppressed at tree-level.
Thus shifting those powers by (at least) $+1$.
Furthermore, $\Nf=12$ is expected to be (near-)conformal for $N=3$ making continuum extrapolations challenging.

\acknowledgments
I am indebted to Gregorio Herdoíza for very helpful suggestions and discussions on the project as well as comments on the manuscript.
I also thank Stefan Sint, Oliver B\"ar, Maarten Golterman, and Agostino Patella for discussions regarding the correct choice of symmetry constraints on the operator basis of the SymEFT action.

The author acknowledges funding by the STFC consolidated grant ST/T000775/1 as well as support of the projects PID2021-127526NB-I00, funded by MCIN/AEI/10.13039/501100011033 and by FEDER EU, IFT Centro de Excelencia Severo Ochoa No CEX2020-001007-S, funded by MCIN/AEI/10.13039/501100011033, H2020-MSCAITN-2018-813942 (EuroPLEx), under grant agreement No. 813942, and the EU Horizon 2020 research and innovation programme, STRONG-2020 project, under grant agreement No. 824093.

%\bibliography{staggered.bib}
%\bibliographystyle{JHEP}

\providecommand{\href}[2]{#2}\begingroup\raggedright\endgroup

\end{document}